
%
\input phyzzx
\Pubnum={UVA-HET-92-01\cr
hepth@xxx/9204013\cr
Revised Version}
\date={April 1992}
\titlepage
\title{Matrix Models and One Dimensional Open String Theory\footnote*
{Supported in part by the United States
Department of Energy under grant DE-AS05-85ER-40518
}}
\bigskip
\author {Joseph~A.~Minahan\footnote\dag
{Electronic mail:  MINAHAN@gomez.phys.virginia.edu}}
\address{Department of Physics,  Jesse Beams Laboratory,\break
University of Virginia, Charlottesville, VA 22901 USA}
\bigskip
\abstract{We propose a random matrix model as a representation
for $D=1$ open strings.   We
show that the model is equivalent to $N$ fermions with spin in a
central potential that also includes a long-range ferromagnetic interaction
between the fermions that falls
off as $1/(r_{ij})^2$.  We find two interesting scaling limits
and calculate the free energy for both situations.  One
limit corresponds to Dirichlet boundary conditions for the dual graphs
and the other corresponds to Neumann conditions.
We compute the
boundary cosmological constant and show that it is of order $1/\log(\beta)$.
We also briefly discuss a possible analog of the Das-Jevicki field for
the open string tachyon.}

\endpage

\def\NP{{\it Nucl. Phys.\ }}
\def\PL{{\it Phys. Lett.\ }}

\def\CMP{{\it Comm. Math. Phys.\ }}

\def\IJMP{{\it Int. Jour. Mod. Phys. A\ }}

\REF\planar{E.~Br\'ezin, C. Itzykson, G. Parisi and J.~B.~Zuber, \CMP
{\bf 59} (1978), 35.}
\REF\KM{V.~A.~Kazakov and A.~A.~Migdal, \NP {\bf B311} (1989) 171.}
\REF\GrMil{D.~J.~Gross and N.~Miljkovi\'c, \PL {\bf B238} (1990), 217.}
\REF\BKZ{E. Br\'ezin, V.~A.~Kazakov and Al.~B. Zamolodchikov, \NP
{\bf B338} (1990), 673.}
\REF\GZJ{P.~Ginsparg and J.~Zinn-Justin, Harvard Preprint, HUTP-90-A004,
1990.}
\REF\DJ{S.~Das and A.~Jevicki, \IJMP {\bf 5} (1990), 1639.}
\REF\SW{A.~Sengupta and A.~Wadia, \IJMP {\bf6} (1991), 1961.}
\REF\Polch{J.~Polchinski, \NP {\bf B362} (1991), 125.}
\REF\MS{G.~Moore and N.~Seiberg, Rutgers Preprint RU-91-29, 1991.}
\REF\Boul{D.~V.~Boulatov, \IJMP {\bf 6} (1991), 79.}
\REF\Kaz{V.~A.~Kazakov, \PL {\bf B237} (1990), 212.}
\REF\Kostov{I.~K.~Kostov, \PL {\bf 238B} (1990), 181.}
\REF\Minahan{J.~A.~Minahan, \PL {\bf B268} (1991), 29.}
\REF\Yang{Z.~Yang, \PL {\bf B257} (1991), 40.}
\REF\Trisurf{V.~A.~Kazakov, \PL {\bf B150} (1985), 282; J.~Ambj\o rn,
B.~Durhuus and J.~Fr\"olich, \NP {\bf B257} (1985), 433;
F.~David, \NP {\bf B257} (1985), 45;
V.~Kazakov, I.~Kostov and A.~Migdal, \PL {B157} (1985) 295.}
\REF\DGK{M.~R.~Douglas, D.~J.~Gross and I.~R.~Klebanov, private communication.}
\REF\GKII{D.~J.~Gross and I.~R.~Klebanov, \NP {\bf B354} (1991), 459.}
\REF\Igor{I.~R.~Klebanov, unpublished.}
\REF\GK{D.~J.~Gross and I.~R.~Klebanov, \NP {\bf B344} (1990), 475.}
\REF\Calog{F.~Calogero, Jour. of Math. Phys. {\bf 12} (1971), 419.}


\def\half{{1\over2}}

\def\al{\alpha}
\def\be{\beta}
\def\gam{\gamma}
\def\alp{\al'}
\def\sralp{\sqrt{\alp}}
\def\psid{\psi^\dagger}

\def\surf{\mit\Sigma}
\def\sF{{\cal F}}
\def\sD{{\cal D}}
\def\sP{{\cal P}}
\def\Lam{\Lambda}
\def\chid{\chi^\dagger}
\def\Ud{U^\dagger}
\def\lam{\lambda}
\def\vm{\Delta}
\def\Seff{S_{\rm eff}}
\def\dt{\Delta t}
\def\ddlam{{\partial^2\over \partial\lam_a^2}}

\def\gob{{\gam\over\be}}
\def\fe{\varepsilon_{\rm F}}
\def\fet{\widetilde\varepsilon_{\rm F}}
\def\muB{\mu_{\rm B}}
\def\muR{\mu_{\rm R}}
\def\kth{k^{\rm th}}
\def\gcs{g_{\rm cs}}
\def\akg{\alp\kappa\gcs}
\def\Jp{J_+}
\def\Jm{J_-}
\def\Jz{J_z}
\def\Jpn{J_+^n}
\def\Jmn{J_-^n}
\def\Jzn{J_z^n}
\def\Spa{S_{+a}}
\def\Sma{S_{-a}}
\def\Sza{S_{za}}
\def\Jv{\vec J}
\def\Jvn{\vec J^n}
\def\Sva{\vec S_a}
\def\Svb{\vec S_b}
\def\gs{|\downarrow\rangle}
\def\dtl{\Delta t_{\ell}}
\def\bd{\cal B}

Matrix models have lead to many insights in low dimensional string theory.  The
model that is perhaps the simplest to understand, yet ironically has the
richest
structure, is that which corresponds to a scalar field theory on a line coupled
to gravity.  This is the matrix quantum mechanics model~[\planar]
with the matrix potential tuned such that the theory is almost
critical~[\KM,\GrMil-\GZJ].
This model is equivalent to a theory of closed strings
in $1+1$ dimensions, where
time is one coordinate and the
eigenvalues of the matrices $\lam$, are related to
the Liouville coordinate~[\DJ-\MS].

A natural extension of this work is to find the corresponding
theory of open and closed strings.   One suggestion is to add the term
$\xi\Tr\log(\phi(t)-\mu)$ to the matrix potential, and tune $\xi$ and
$\mu$ to some appropriate values~[\Boul].
The trouble with this is that, unlike
the zero dimensional case~[\Kaz-\Minahan],
this term does not arise  by integrating
out fields which are $1\times N$ matrices that couple to $\phi$.
These fields are fundamental representations of a global $U(N)$ group and
it is precisely these fields that generate boundary terms in the
Feynman diagrams.  The problem with the log term is that the kinetic piece
of the fundamental fields has been ignored.  One might argue that it is all
right to drop this term
if the mass and the couplings are very large.
However, it has been pointed out that the interesting critical behavior occurs
when the argument of the log approaches zero~[\Boul,\Yang].
This is precisely where the kinetic piece should become important.

In this paper we explicitly include these fundamental fields
in the full lagrangian.
We demonstrate that the theory is equivalent to $N$ fermions with spin
which interact among each other with long-range ferromagnetic interactions.
We show that there are two interesting
scaling limits for the masses of the fundamental fields.
One limit has the mass
diverging logarithmically as $N\to\infty$, while the other case has a mass
that diverges, but at a milder rate.  The former case appears to correspond
to an open string theory with Dirichlet boundary conditions on the dual
graphs
(and hence Neumann conditions on the original graphs),
while the latter limit corresponds to a theory with
Neumann conditions.
In both cases we compute the free energy as
an expansion in the open and closed string couplings.  The Neumann case
has almost all terms in the expansion negative definite, while the
Dirichlet case has the same form for the free energy,
but with the opposite sign for the open string coupling.
The Neumann case also has a critical value for the open string coupling, at
which point divergences start appearing in the free energy.
We also show that the boundary cosmological constant scales as
$(\log \be)^{-1}$.
Finally, we discuss a possible analog to the Das-Jevicki collective coordinate
field~[\DJ] for the open string tachyon.

To begin,
consider a triangulated surface $\surf$, with boundary $\bd$, where $\bd$
is not necessarily connected.  Suppose that $\surf$ is the world-sheet for a
string theory which has a coordinate $t$.  Each vertex of a triangle marks
a particular point in $t$ space, $t_v$. Then an edge, $\ell$, connecting
two vertices has associated with it a difference in $t$, $\dtl=t_{v_1}-
t_{v_2}$.  The complete partition function should be comprised of a sum
over all assignments of $t_v$ to the vertices.  This is equivalent to summing
over all $\dtl$, so long as we impose the constraint
$$\sum_{\ell\in C}\dtl=0,\eqn\Constraint$$
where the sum over $\ell$ is over the three edges of any triangle $C$.

The string world sheets should also have boundary conditions.  For the
triangulated surface $\surf$, Neumann boundary conditions impose the constraint
$\dtl=0$ if $\ell$ is an edge that intersects the boundary.
On the other hand, Dirichlet boundary conditions impose the same constraint
$\dtl=0$, but now $\ell$ is an edge that lives on the boundary.

The dual surface of $\surf$ is found by bisecting all edges.  This operation
maps vertices to faces and {\it vice versa}, while edges are mapped to
new edges at 90 degrees.
The dual surface can be generated by a field theory of $N\times
N$ matrices, $\phi_{ab}$, with cubic interactions~[\Trisurf,\KM].
The surfaces are the Feynman diagrams for this theory, with the
$\phi$ propagators forming the edges.
$\dtl$ for an edge in $\surf$ maps to $p_l$, the momentum flowing through
the bisecting propagator,
while the constraint in \Constraint\
naturally maps to the constraint that the sum of the momenta entering any
vertex is zero.

The boundaries of $\surf$ present a slight problem when finding the dual graph.
The edge dual to a boundary edge has nothing to attach to outside
the boundary.  We can remedy this by introducing new fields, $\psi_a$ and
$\chi_a$ which transform in the fundamental representation of a global
$U(N)$ and which couple to $\phi_{ab}$.  The propagators of these
fields can then be used to tie off the ends of the $\phi$ propagators
dangling over the boundary.  $\psi_a$ and $\chi_a$ are assumed to be
fermionic and we will see later that two fields are necessary
in order to have a nontrivial theory.

The boundary conditions in $\surf$ lead to analogous constraints for the
dual graphs~[\DGK].   Edges that are normal to the boundary are mapped to edges
that are parallel.  Therefore, Neumann conditions on $\surf$ correspond
to dual graphs with zero
momentum in the $\phi$ propagators that run along the boundary.
This means that if a large amount of momentum flows into a region of
the Feynman diagram which includes a long stretch of the boundary,
then the momentum should mainly flow through the propagators
of the fundamental field and not the $\phi$ propagators that are next to
but not actually intersecting
the $\psi$ or $\chi$ propagators.  This is accomplished
by making the masses of $\psi_a$ and $\chi_a$ very large,
in which case, the weights of the Feynman diagrams strongly favor
a large amount of momentum flowing through the $\psi$ and $\chi$
propagators.
As the masses diverge, the correlation lengths of these fields shrink to
zero and all fermion loops become localized in $t$.
Thus,
Neumann conditions on $\surf$ lead to Dirichlet conditions for the
dual theory.  Likewise, Dirichlet conditions on $\surf$ lead to dual
graphs with zero momenturm flowing off the boundaries.
Such graphs
will dominate the free energy if the masses of $\psi_a$ and $\chi_a$ are
chosen to be small.  The dual graph then has Neumann boundary
conditions.

Let us consider the action, previously considered by Yang~[\Yang]
$$\eqalign{S=\be\int dt\biggl\{
\Tr \Bigl(\half\dot\phi^2-V(\phi)\Bigr)&+i\psid_a\dot\psi_a
+i\chid_a\dot\chi_a\cr
&+\gam\psid_a\phi^2_{ab}\psi_b+\gam\chid_a\phi^{*2}_{ab}\chi_b
-\mu\psid_a\psi_a-\mu\chid_a\chi_a\biggr\}.}\eqn\action$$
We have chosen to couple the fermion fields to the square of the hermitian
matrix for later convenience, but this won't affect the critical
behavior in the double scaling limit.
It is necessary to couple $\chi$ to the complex conjugate of $\phi$
in order to build open string states
that are invariant under a global $SU(N)$ rotation.

In the large $N$ limit the free energy derived
from this action is dominated by the planar diagrams and
can be expressed as an expansion in $1/N$,
$$F=-\sum_{\rm surfaces} N^{2(1-g)}N^{-h}(N/\beta)^{\rm area}
(\gam^2N/\beta)^{\rm length}\sF(\surf,g,h,\mu),\eqn\surfamp$$
where $g$ is the genus of the surface and $h$ is the number of holes.
Each vertex in the dual diagram is weighted by $(N/\be)^{1/2}$ and each vertex
on the boundary is weighted by $\gam(N/\be)^{1/2}$.  Hence in terms of the
original triangulated graph, these weights
lead to the terms in the free energy
exponentiated by the area and length in appropriate units.
$\sF(\surf,g,h,\mu)$
is a weight for the Feynman diagram corresponding to a
particular surface.
If $\mu$ is very large then the $\mu$ dependence of $\sF$ is approximately
$(1/\mu)^{\rm length}$.

To proceed, let us diagonalize the matrix $\phi$,
$\phi=\Ud\Lam U$,
where $\Lam$ is diagonal and $ U$ is an element of $SU(N)$.  Moreover,
let us rotate the fermion fields $\psi_a\to\Ud_{ab}\psi_b$ and
$\chi_a\to\chi_b U_{ba}$.  Then the action
\action\ becomes
$$\eqalign{S=\be&\int dt  \sum_a\Bigl(\half\dot\lam^2_a-V(\lam_a)
+i\psid_a\dot\psi_a +i\chid_a\dot\chi_a\cr
&\qquad\qquad
+\gam\psid_a\lam_a^2\psi_a+\gam\chid_a\lam_a^2\chi_a
-\mu\psid_a\psi_a-\mu\chid_a\chi_a\Bigr)\cr
&-\sum_{a\ne b}
\Bigl(\half(\dot U\Ud)_{ab}(\lam_a-\lam_b)^2(\dot U\Ud)_{ba}
+i\psid_a(\dot U\Ud)_{ab}\psi_b
-i(\dot U\Ud)_{ab}\chid_b\chi_a\Bigr).}\eqn\actiond$$
Letting $A=i\dot U\Ud$, we see that the action is quadratic in $A$ and
hence can be integrated out after shifting variables~[\GKII,\Igor].
To this end, consider the path integral for propagation of the vacuum state
from time $t=0$ to time $t=T$,
$$\eqalign{Z=&\int\sD \phi\sD\psi\sD\psid\sD\chid\sD\chi e^{iS}\cr
=&\int\sD\lam\sD U\sD\psid\sD\psi\sD\chid\sD\chi
\prod_t(\vm(\lam(t))^2 e^{iS},}\eqn\pathint$$
where $\vm(\lam(t))$ is the vandermonde determinant
$$\vm(\lam(t))=\prod_{a<b} (\lam_a(t)-\lam_b(t))^2.\eqn\vmeq$$
Completing the square and shifting
$A$, leads to the extra term in the action~[\Igor]
$$-\be\int dt \sum_{a\ne b}{(\psid_a\psi_b-\chid_b\chi_a)
(\psid_b\psi_a-\chid_a\chi_b)\over
(\lam_a-\lam_b)^2}.\eqn\effaction$$
Noting that $\Ud(t+\dt) U(t)=1+iA(t+\dt/2)\dt+...$, we can replace the $ U$
integrals with $A$ integrals,  less one $U$ integration because of
the global $SU(N)$ symmetry.
Since the diagonal elements of $SU(N)$ commute with $\Lam$, we should
restrict the integration to a subspace of $A$.  If we only consider the
insertions of $SU(N)$ invariant states into the path integral, then this
subspace is just the off-diagonal elements of $A$.
This integration over $A_{ab}$ leads to factors that cancel
off the vandermonde determinants in the Jacobian,
except for a leftover contribution at each end point
because of the extra integration over an $SU(N)$ variable.
Hence the partition function reduces to
$$Z=\int\sD\lam\psid\sD\psi\sD\chid\sD\chi \vm(\lam(0))\vm(\lam(T))
 e^{i\Seff},\eqn\effpathint$$
where $\Seff$ is the effective action after integrating out $A$.

The fermion kinetic terms lead to the standard quantization for $\psi_a$ and
$\chi_a$,
$$\{\psi_a,\psid_b\}=\delta_{ab},\qquad\qquad
\{\chi_a,\chid_b\}=\delta_{ab},\eqn\fermcomm$$
where a factor of $\be^{1/2}$ has been absorbed into the fermion fields.
Therefore, the hamiltonian for $SU(N)$ invariant states is given by
$$\eqalign{H=\sum_a\biggl(-{1\over2\be^2}\ddlam+V(\lam_a)&+{\gam\over\be}
(\psi_a\lam_a^2 \psid_a+\chi_a\lam_a^2\chid_a)
-{\mu\over\be}(\psi_a\psid_a+\chi_a\chid_a)\biggr)\cr
&-{1\over\be^2} \sum_{a< b}
{(\psid_a\psi_b -\chid_b\chi_a) (\psid_b\psi_a -\chid_a\chi_b)
\over (\lam_a-\lam_b)^2}.}\eqn\Ham$$
The last term in the hamiltonian leads to a repulsive force between the
eigenvalues.

However, because $\psi_a$ and $\chi_a$ don't commute with $\psid_a$
and $\chid_a$ respectively, it is
necessary to determine
the ordering of the operators in $H$ so that the theory is
well defined.   The ordering follows from the definition of the time derivative
in the path integral.  As is the case with the kinetic term for the hermitian
matrix, the kinetic term for the fermions should link an otherwise uncoupled
chain of fields.  With this in mind,
let us define the kinetic term for $\psi_a$ as
$$\sum_{t}i\psid_a(t)(\psi_a(t)-\psi_a(t-\dt)).\eqn\kinetic$$
Rotating $\psi_a$ by $\Ud$ and
integrating out the angular variables leaves the following contribution to
the path integral (ignoring the $\chi$ field):
$$\eqalign{Z_{\psi}(\lam)=\int\sD\psid\sD\psi \exp\biggl\{
&i\sum_{t=0}^{t=T}\Bigl[
\sum_a i\psid_a(t)\bigl(\psi_a(t)-\psi(t-\dt)\bigr)\cr
&+\dt\bigl(\gam\psid_a(t)\lam_a^2(t) \psi_a(t)-\mu\psid_a(t)\psi_a(t)\bigr)
\Bigr]\cr
&+\sum_{a<b}{\dt\over\be}{\psid_a(t)\psid_b(t)
\psi_b(t-\dt)\psi_a(t-\dt)\over (\lam_a-\lam_b)^2}\biggr\}}\eqn\fermpi$$
where the integrations over the $\psi$ variables are at the points between
$t=0$ and $t=T$ inclusive.  Then, with no $\psi$ insertions in
$Z_\psi(\lam)$, \fermpi\ reduces to
$$\eqalign{Z_\psi(\lam)=&C\prod_t\Bigl(1-i{\gam}\dt
\sum_a(\lam^2_a(t)-\mu)\Bigr)\cr
=&C\exp\Bigl(-i\be\int_0^T dt\gob\sum_a(\lam^2_a(t)-\mu)\Bigr),}
\eqn\fermpii$$
where $C$ is an unimportant constant.
If we now insert the $SU(N)$ invariant operator
$\epsilon_{a_1a_2...a_N}\psid_{a_1}\psid_{a_2}
 ...\psid_{a_N}$ into $Z_{\psi(\lam)}$ at $t=0$,
we find that the contribution to the path integral is $C$.
This means that there is no contribution to the energy from the four fermi
interaction if the $\psi$ states are either completely full or empty.
Thus we learn that the explicit
ordering given by the hamiltonian in
\Ham\ follows from our definition of the infinite matrix chain.

{}From the hamiltonian \Ham\ we can determine the ground state, which should
be a state that is $SU(N)$ invariant.  Given an $SU(N)$ invariant
state $|\Psi\rangle$,
other invariant states can be constructed by acting
on $|\Psi\rangle$ with not only the operator $\tr\phi^n$, but also with
$\Jpn$, $\Jmn$ and $\Jzn$, where these last three operators are defined as
$$\eqalign{\Jpn&=\psid_a\phi^n_{ab}\chid_b,\qquad\qquad
\Jmn=\chi_a\phi^n_{ab}\psi_b,\cr
\Jzn&=\half(\psid_a\phi^n_{ab}\psi_b-\chi_a\phi^n_{ab}\chid_b).}\eqn\Jops$$
The $J^n$ operators form an algebra
$$[\Jz^n,J_{\pm}^m]=\pm J_{\pm}^{n+m}\qquad\qquad
[\Jp^n,\Jm^m]=2\Jz^{n+m},\eqn\Jalg$$
and in particular, the operators $\Jv\equiv\Jv^0$ are the generators
of an $SU(2)$ lie algebra.
Furthermore, after diagonalizing $\phi$ and redefining $\psi$ and $\chi$
we can construct operators $\Spa$, $\Sma$ and $\Sza$, where
$$\Spa=\psid_a\chid_a,\qquad\qquad\Sma=\chi_a\psi_a\qquad\qquad
\Sza=\half(\psid_a\psi_a-\chi_a\chid_a),\eqn\Sops$$
and which satisfy the relation $\Jv^n=\sum_a\lam_a^n\Sva$.  Clearly,
for each index $a$, $\Sva$ can be interpreted as a spin operator.
Hence the hamiltonian \Ham\ can be re\"expressed as
$$H=\sum_a\biggl(-{1\over2\be^2}\ddlam+V(\lam_a)-{2\over\be}
(\Sza-\half)(\gam\lam_a^2-\mu)\biggr)
+{2\over\be^2} \sum_{a< b}
{1/4-\Sva\cdot\Svb \over (\lam_a-\lam_b)^2}.\eqn\Hamnew$$
The result is a system equivalent to $N$ fermions with spin
in a central potential which includes a  position dependent
magnetic field and with Heisenberg
ferromagnetic long range couplings between the spins.
Although $\Jv\cdot\Jv$ does not commute with $H$, $\Jz$ does,
so the energy eigenstates can be classified by their $\Jz$ quantum numbers.

It is not known how to calculate the exact spectrum of \Hamnew.  However,
for some choices of $\gam$ and $\mu$ we can at least find the ground state.
First consider the case $\gam<0$.  If $\mu\ge0$, then the magnetic
field, while position dependent, points in the same direction for
all values of $\lam_a$.  Hence, the ground state $\gs$,
has all spins pointing down
in the direction of the field and $\Jz=-N/2$.
The reduced hamiltonian for spins with this configuration
is~[\Yang]
$$H_{\downarrow}=\sum_a\biggl(-{1\over2\be^2}\ddlam+V(\lam_a)+{2\over\be}
(\gam\lam_a^2-\mu)\biggr).\eqn\Hamred$$
Since all spins are down, the fermion interaction term has dropped out.
If we now let $\mu<0$, then $\gs$ is not necessarily the ground state
anymore, since the field is no longer pointing in the same direction for
all $\lambda$.  For some negative
value of $\mu$, the lowest energy state with
$\Jz=-N/2$ becomes degenerate with the lowest energy state with $\Jz=1-N/2$.
Hence, this value of $\mu$ is a critical point for the fermion mass.

Next consider the case $\gam>0$.
If $\mu<0$, then the field is pointing in the same direction
for all $\lam$, but now the ground state has all spins up.  Examining
the hamiltonian \Hamnew, we find that there is no longer any $\gam$ dependence
in the reduced hamiltonian.  However,
if we assume that $V(\lam)$ is infinite in the range
$|\lam|>a$,
and if $\mu>\gam a^2$, then the field points in the same direction for all
$|\lam|<a$ and the ground state is spin down.
In this case, the reduced hamiltonian \Hamred\ is valid, and there
exists a critical value of $\mu$,
$\mu_0\approx\gam a^2$

The net effect of the $\psi$ and $\chi$ fields on the ground state
is to shift the potential for $\lam$, and
with proper scaling of $\gam$, to shift the fermi energy.
Let the potential $V(\lam)$ be given by
$$\eqalign{V(\lam)=&\bigl(\lam^2-\lam^4/(2a^2)\bigr)/\alp\qquad\qquad
-a\le\lam\le a\cr
=&\infty\qquad\qquad\qquad\qquad\qquad\qquad|\lam|>a}\eqn\quartpot$$
where $\alp$ is the Regge slope.
This potential has maxima at $\lam=\pm a$, and at these points
$V''(\pm a)=-1/\alp$.
We have chosen the infinite wall to lie at the local maxima so that we can
have a small magnetic field near these points and still be able to
compute the ground state.
Adding the term in \Hamred\ shifts the maxima to
$\pm a(1+2\gam\alp/\be)^{1/2}$
and changes the second derivatives at these points to
$-(1+2\gam\alp/\be)/\alp$.
Furthermore, if $\mu$ is tuned close to $\mu_0$ then the magnetic field is
small near the local maxima.\foot{
We should note that for $\gam>0$, the points $\pm a(1+2\gam\alp/\be)^{1/2}$
are in the region where the potential is infinite.  But this won't matter
because, as we will see later, $a\gam\alp/\be$ is much smaller than
$((V(a)-\fe)/\alp)^{1/2}$.  This is the distance from the wall where
the potential equals the fermi energy.}

The quartic potential is rather unwieldy to work with, so instead
let us use the potential
$$\eqalign{V(\lam)=&-\lam^2/\alp\qquad\qquad0\le\lam\le a\cr
=&\infty\qquad\qquad\lam<0\qquad{\rm or}\qquad\lam>a.}\eqn\quadpot$$
Now the effect of the boundary fields is to multiply $V(\lam)$ by
$(1+2\gam\alp/\be)$.
The fermi energy for the  potential in \quadpot\ is easily found using a
semiclassical approximation for the density of states~[\KM].
Inserting this approximation into an integral over phase space
gives an expression for $N$,
$$N=\be\int_{0}^a d\lam \int{dp\over2\pi}
\theta(\fe-p^2/2+\lam^2/\alp).\eqn\Ndens$$
The integral is straightforward, giving the relation
$$N={\sqrt{2}\be(-\alp\fe)\over2\pi\sqrt{\alp}}\biggl(
{a\over\sqrt{-\alp\fe}}\sqrt{{a^2\over-\alp\fe}-1}-\cosh^{-1}
(a/\sqrt{-\alp\fe}).
\biggr)\eqn\Ndensii$$
For our purposes we can approximate this equation as
$$N={\sqrt{2}\be\over2\pi\sralp}\Bigl(a^2
+{1\over4}(-\alp\fe)\log(-\fe)+{\rm O}(\fe)\Bigr).
\eqn\Napprox$$

Shifting the potential leads to a new fermi energy $\fet$,
which satisfies the equation
$$N={\sqrt{2}\be\over2\pi\sqrt{\alp}}\Bigl(a^2\sqrt{1+2\gam\alp/\be}+
{1\over4}(-\alp\fet)\log(-\fet)+{\rm O}(\fet)\Bigr).
\eqn\Nnapprox$$
The value of $\mu$ does not effect this expression so long as
the ground state remains spin down.
Equating the expressions for $N$ in \Napprox\ and \Nnapprox\
leads to the relation
$$\gam={\be\over4a^2}(\fet-\fe)\log(-\fe).\eqn\gameq$$
It has been argued~[\BKZ,\GZJ,\GK] that the proper
double scaling limit is found by fixing
$\be\sralp\fe$ to be constant while taking
the limit $N\to\infty$ and $\be/N$ to its critical value.
Thus, we discover that $\gam$ must have a logarithmic divergence in the scaling
limit in order to shift $\be\fe$ by a finite amount.

In order to understand the significance of this shift, let us consider the
free energy in the scaling limit.  Many authors have argued that
this free energy is given by~[\GrMil,\BKZ,\GZJ,\GK]
$$\eqalign{F={1\over4\pi\be\sralp}
\biggl(-(\be\sralp\fe)^2&\log(-\fe)+{1\over3}\log(-\fe)\cr
&-\sum_{m=1}^{\infty}(2^{2m+1}-1){|B_{2m+2}|\over m(m+1)}(\be\sralp\fe)^{-2m}
\biggr),}\eqn\freeenergy$$
where $B_{2m}$ are the Bernoulli numbers.  Based on this expansion,
it seems clear that $(-\be\sralp\fe)^{-1}$ should be interpreted as the
closed string coupling, $\gcs$.  If we shift the fermi energy to $\fet$ and
define $\kappa=\be(\fet-\fe)/\sralp$, then the relevant part of
the free energy becomes
$$\eqalign{F={1\over4\pi\be\sralp}\biggl(-{1\over \gcs^2}&
\Bigl\{(1-\akg)^2\log(-\fe)-\akg\cr
&+{3\over2}(\akg)^2
-2\sum_{n=2}^\infty{
(\akg)^n\over n(n-1)(n-2)}
\Bigr\}\cr
&+{1\over3}\log(-\fe)-{1\over3}\sum_{n=1}^\infty
(\akg)^n\cr 
&-\sum_{m=1\atop n=0}^{\infty}(2^{2m+1}-1){|B_{2m+2}|\over m(m+1)}\gcs^{2m}
{
(\akg)^n(2m+n)!\over(2m)!}
\biggr),}\eqn\nfreeenergy$$
Shifting the bulk cosmological constant shifts $\fe$ and $\fet$ by an equal
amount and thus keeps $\kappa$ fixed while scaling $\gcs$.  A surface
of genus $g$ with $h$ holes should scale as $(\gcs)^{2(1-g)+h}$.  Hence
the terms in the expansion of the free energy have the proper scaling if
$\kappa$ is interpreted as an open string coupling constant.

If $\gam<0$, then $\fet>\fe$ and $\kappa>0$.
In this case all terms in the expansion of the free energy are negative
definite,
except for those terms inside the curly brackets in \nfreeenergy.
{}From \nfreeenergy\ we also see that $\kappa=1/(\gcs\sralp)$ is a critical
value for the coupling constant since $\fet=0$ for this value, and thus
the free energy for surfaces of genus
$g\ge1$ diverges when summing over the number of holes.
The significance of this is not presently clear.

Letting $\gam>0$ means that
$\kappa<0$ and therefore the terms in \nfreeenergy\ alternate sign.
Furthermore, $\mu$ must satisfy
$$\mu>\mu_0\approx\gam a^2=-(\kappa\sralp/4) \log(-\fe)\eqn\mucond$$
in order that the ground state is spin down.
Therefore, in the scaling limit, the mass of the boundary generating fields
diverges and hence,
the dual graphs have Dirichlet boundary conditions.

We can give a rough estimate of the correction to $\mu_0$ coming from the
ferromagnetic interaction.  If we flip one spin up, then the
ferromagnetic term will try to symmetrize the up spin among all the fermions.
This is counteracted by the position dependent field which tries to push the
spin to a point where the field is small.  We will approximate these effects
by assuming that the up spin is symmetrized among the fermions less than a
distance $2y$ from the top of the potential at $\lam=0$.  Hence, we will
say that the average position of the
up spin is at $\lam=y$ and that it feels the $1/r^2$ potential from the down
spins that are greater than a distance $y$ away from it.
Thus, we estimate the shift in energy for turning up one spin as
$$\Delta E\sim {1\over\be^2}\int_{y}d\lam {\rho(\lam)\over \lam^2}
-{2\Delta\mu_0\over\be}+{2\over\be}\gam y^2\eqn\Eshift$$
where $\rho(\lam)$ is the density of states, $dn\over d\lam$.
Plugging in the expression for
$\gam$ in \gameq\ and using the semiclassical estimate
$$\rho(\lam)={\sqrt{2}\be\over\pi\sralp}\sqrt{\lam^2+\fet\alp}\eqn\rhoeq$$
gives
$$\Delta E\sim {\sqrt{2}\over\pi\be\sralp}\log y+
{\sralp\kappa\log(-\fe)\over\be a^2}y^2-{2\Delta\mu_0\over\be}.\eqn\EshiftII$$
Minimizing $\Delta E$ leads to the estimates
$$y\sim {a\over(\alp\kappa\log(-\fe))^{1/2}},\qquad
\Delta E\sim
(\sralp\be)^{-1}\bigl(\log(\alp\kappa\log(-\fe))-\Delta\mu_0\bigr).
\eqn\esteq$$
Therefore the correction to $\mu_0$ is of order
$$\Delta\mu_0\sim\log(\alp\kappa\log(-\fe)).\eqn\mucorr$$
Actually, given the roughness
of our estimate, it is not clear that we can distinguish between a double
log and a constant.  In particular, flipping the spin will certainly alter
the density of states, and if the spin pushes the eigenvalues
out far enough, then the double log should be replaced with
a constant.  In any case the correction $\mu_0$ is much smaller than $\mu_0$

Now suppose that $\gam<0$ and $\mu\approx0$.
Let us calculate the energy to flip one spin at the bottom of the quartic
potential.
We can play the same game to estimate the critical value of $\mu$.
In this case turning up one spin shifts the energy by
$$\Delta E\sim{2\sqrt{2}\over\be\pi\sralp}\int_y dx
{\sqrt{x^2+a^2}\over x^2} +{2\mu_0\over\be}
-{\kappa\sralp\log(-\fe)\over 2a^2}y^2.\eqn\Eshiftn$$
This then leads to the estimate
$$\mu_0\sim-(a\be)^{-1}(-\alp\kappa\log(-\fe))^{1/3}\eqn\mucorrn$$
in order to keep $\Delta E$ small.
Hence in order to have some relevance in the double scaling limit,
the Neumann case also requires a divergent mass for the fundamental fields,
although the divergence is milder than in the Dirichlet case.
In fact, the mass is small in the sense that
$\mu_0/\gam\to0$ in the double scaling limit.

We would like to calculate the change in energy to turn up one spin
if there are already $s$ spins pointing up.  This then leads to
a density of up spins $ds\over dE$.  It is not clear how to do this, but
a reasonable guess for this behavior is that
the density scales as $\be\log(-\fe)$ for the Dirichlet case and
$\be$ for the Neumann case.  We postulate that
the extra log term in the Dirichlet case
is a result of the magnetic field being small at the top of the potential.
Near this point, while there are fewer fermions around whose spin can flip,
there are also fewer fermions to oppose such a flip.  The guess is that
the latter effect dominates the former.

Turning to the one point functions of $\Jvn$, it is clear that only
$\langle\Jzn\rangle$ is nontrivial.  Actually, based on the structure
of the hamiltonian in \Hamnew, the relevant operator is $(1/N)
(\tr\phi^n-2\Jzn)$,
which couples to the fermion loops.   Because of the structure of the
vacuum, the one-point function for this
operator is the same as the one-point function for $(2/ N)\tr\phi^n$.
Note that if $\mu>>\mu_0$ then
those states that are constructed by acting on the vacuum with $\Jpn$
decouple from the theory.  In this case, the theory is essentially equivalent
to a theory containing only closed strings, since now
all $\Jz^n$ operators can be replaced by $\tr\phi^n$ operators.
The only effect of $\psi$ and $\chi$ in this case is to
renormalize the closed string coupling constant.

One interesting operator is $\sP=1-2\Jz^0/N$, where $\sP$ generates
linear transformations of $\mu$.  If
$\mu\ge\mu_0$, then the one
point function for $\sP$ is
$$\langle\sP\rangle=2.\eqn\onepoint$$
In the case of dual graphs with Dirichlet boundary conditions, the
length dependence of the free energy
for a particular surface is $(\gam/\mu)^{\rm length}$.
Therefore, $\sP$ generates a shift of the boundary cosmological constant.
For $\mu\gsim\mu_0$, the boundary cosmological constant is, in some
appropriate units,
$$\muB=(\mu-\mu_0)/\mu_0={4(\mu-\mu_0)\over\kappa\sralp\log(-\fe)}.\eqn
\boundcc$$
As in the case for closed strings, the cosmological constant should
approach zero as the system nears the critical point.
Since $\sP$ generates transformations of order unity,
the value of $\mu-\mu_0$ should also be of order unity
in order that it have some relevance in the scaling limit.
In other words, if $\mu$ is shifted by order unity, then the energies of
states with one spin pointing up are shifted by order $1/\be$, which is enough
to have some effect on the physics.
Thus, we find that $\muB$ is of order $(-\log(-\fe))^{-1}$.

It is instructive to compare this with the behavior of the boundary
cosmological constant for the one-matrix model.
For the $\kth$ multicritical point, the bare
boundary cosmological constant scales as $\muB=\be^{-2/(2k+1)}\muR$,
where $\muR$ is the renormalized value~[\Kostov,\Minahan].
As $k\to\infty$, $c$ approaches
$1$ and the $\be$ dependence of $\muB$ is $\be^\epsilon$ with $\epsilon\to0$.
This result compares favorably with our result for the one-dimensional chain.

The open string vertex operators should be constructed out of some combination
of the $\Jvn$ operators, which raise and lower the spins of the fermions.
This suggests a possible connection to
the Das-Jevicki collective field formulation of the tachyon field~[\DJ].
Recall that these authors showed there was a massless field associated with
the fluctuations of the eigenvalue density, which they identified with the
closed string tachyon.  Analagously, it seems that the
open string tachyon should be
associated with the density of up pointing spins.  This means that in the
classical picture, there should be a connection between propagating tachyons
and spin waves in the matrix model.


In order to compute nontrivial open string scattering amplitudes, it
is necessary to compute the spectra of states that have some
spins up.  It is not clear if it will be possible to do this exactly, but it
is worthwhile to point out that the interaction term
between the fermions in \Hamnew\ is somewhat similar to the
interaction studied by Calogero~[\Calog].   Calogero had the $1/r^2$ potentials
but not the spin-spin interaction.
In any case, he showed that with a harmonic central potential, his
theory was exactly solvable.  It would thus seem not totally pointless
to pursue exact solutions of \Hamnew.  Work on this issue and others
is in progress.

\ack{I would like to thank I. Klebanov, M. Douglas and D. Gross for
very helpful discussions.}

\refout
\end